\documentclass{elsart}

\begin{document}

\begin{frontmatter}



\title{
${\ PT}$ symmetry breaking and explicit expressions for the 
pseudo-norm in the Scarf II potential
}


\author[label1]{G. L\'evai} 
\address[label1]{
Institute of Nuclear Research of the Hungarian 
         Academy of Sciences,
         PO Box 51, H--4001 Debrecen, Hungary}
\ead{levai@atomki.hu}

\author[label2]{F. Cannata} 
\address[label2]{
Dipartimento di Fisica dell'Universit\`a and 
         Istituto Nazionale di Fisica Nucleare, 
         I-40126 Bologna, Italy}

\author[label3]{A. Ventura}
\address[label3]{
Ente Nuove Tecnologie, Energia e Ambiente, 
         Bologna, Italy}

\begin{abstract}
Closed expressions are derived for the pseudo-norm, norm and 
orthogonality relations for arbitrary bound states of the ${\ PT}$ 
symmetric and the Hermitian Scarf II potential for the first time. 
The pseudo-norm is found to have indefinite sign in general. 
Some aspects of the spontaneous breakdown of ${\ PT}$ symmetry 
are analysed. 
\end{abstract}

\end{frontmatter}

\section{Introduction}

Non-Hermitian quantum mechanical problems have attracted much attention 
recently. The main reason for this is that the energy spectrum of a 
number of complex potentials turned out to be real (at least partly), 
which contradicted the usual expectations regarding non-Hermitian 
systems. Strangely enough, the first examples for complex 
potentials with real spectra were found using numerical techniques 
\cite{bb98}. This unusual behaviour of the energy spectrum was 
attributed to the so-called ${\ PT}$ symmetry, i.e. the invariance 
of the Hamiltonian with respect to the simultaneous space (${\ P}$) 
and time (${\ T}$) reflection. For one-dimensional potential problems 
this requires $[V(-x)]^*=V(x)$, which implies that the real component 
of the potential must be an even function of $x$, while the imaginary 
component has to be odd. After the first examples, further ones have been 
identified using semiclassical \cite{semi}, numerical \cite{num} and 
perturbative \cite{pert} methods, and a number of exactly solvable 
${\ PT}$ symmetric potentials have also been found, mainly as 
the analogues of Hermitian (real) potentials \cite{mz,lgmz00,br00}. 

It was also noticed that ${\ PT}$ symmetry is neither a necessary, 
nor a sufficient condition for having real energy spectrum in a 
complex potential. It is not a necessary condition, because there 
are complex non-${\ PT}$ symmetric potentials with these properties: 
some of these are complex supersymmetric partners of real potentials 
\cite{ptsusy}, while some others can be obtained by merely shifting 
${\ PT}$ symmetric potentials along the $x$ axis, which (formally) 
cancels ${\ PT}$ symmetry, but obviously does not influence the energy 
spectrum. Neither is ${\ PT}$ symmetry a sufficient condition, 
because complex-energy solutions of such potentials are also known, 
and since in this case the energy eigenfunctions cease to be 
eigenfunctions of the ${\ PT}$ operator, this scenario has been 
interpreted as the spontaneous breakdown of ${\ PT}$ symmetry 
\cite{bb98}. No general condition has been found for the breakdown 
of ${\ PT}$ symmetry, but it has been observed that it usually 
characterizes strongly non-Hermitian problems 
\cite{bb98,ahmed01a,mzlg01,km00}. 

Obviously, the lack of Hermiticity raises questions about the 
probabilistic interpretation of the wavefunctions (probability density, 
continuity equation), and in general, about the definition of the norm 
and the inner product of the eigenvectors of the non-Hermitian Hamiltonian. 
It has been suggested, for example, that the $\psi^2(x)$ quantity 
should replace $\vert \psi(x)\vert^2$ in the definition of the norm 
\cite{bcms99}. For unbroken ${\ PT}$ symmetry this expression 
coincides with the $\psi(x)\psi^*(-x)$ quantity used in the definition 
of the pseudo-norm \cite{mz0103054}, which is obtained from the 
modified inner product 
$\langle \psi_i \vert {\ P}\vert \psi_j\rangle$. 
This redefinition of the inner product was found to lead to the 
orthogonality of the energy eigenstates, but it also resulted in 
an indefinite metric, replacing the usual Hilbert space with 
the Krein space \cite{japaridze}. Efforts have been made to 
restore the Hermitian formalism using projection techniques 
\cite{mz0103054,japaridze}. 

Exactly solvable examples can be extremely useful in the understanding 
of the unusual features of ${\ PT}$ symmetric problems and the 
underlying new physical concepts. For example, by the continuous tuning 
of the potential parameters through critical values the mechanism of 
${\ PT}$ symmetry breaking can be studied; conditions for positive 
and negative values of the pseudo-norm can be identified, etc. Although a 
number of exactly solvable ${\ PT}$ symmetric potentials have been 
identified, these questions have been addressed only in very few cases. 

The Scarf II (sometimes also called hyperbolic Scarf or Gendenshtein) 
potential is in a special position among exactly solvable potentials.  
This shape-invariant potential \cite{si} is defined on the whole $x$ 
axis, it has no singularity at $x=0$, and in contrast with most other 
shape-invariant potentials, it can be turned into a ${\ PT}$ 
symmetric form without regularizing its singularity by means of an 
$x\rightarrow x+{\rm i}\epsilon$ imaginary coordinate shift 
\cite{lgmz00,ahmed01b}. Therefore it is not surprizing that it became 
a ``guinea pig'' of testing ${\ PT}$ symmetry on a solvable example. 
It has been associated with the sl(2,C) \cite{bbcq00}, 
su(1,1)$\simeq$so(2,1) \cite{lgfcav01} and so(2,2) \cite{lgfcav02} 
potential algebras, and it has also been observed that its ${\ PT}$ 
symmetric version has a second set of bound states, which appear as 
resonances in its Hermitian version \cite{bbcq00,lgfcav01}. 
This mechanism of doubling the bound states 
is essentially different from the one arising from the cancellation 
of singularities at $x=0$ by the imaginary coordinate shift. 
This potential is also known to have (purely) real and 
(purely) complex energy spectrum, depending on the relative strength 
of its real and imaginary component \cite{ahmed01a}, and since 
the two domains can be connected with a continuous tuning of the 
parameters without crossing a singularity, it is a perfect example 
to illustrate the breakdown of ${\ PT}$ symmetry. 

It would also be a suitable example to illustrate the modified 
definition of the inner product and the behaviour (sign) of the 
pseudo-norm, and the other implications for the use of the Krein space 
instead of the usual Hilbert space, however, 
there is a major obstacle: 
the evaluation of integrals involving them could not be calculated as 
yet analitically, except for the ground state \cite{bbcqmz01}. In 
fact, even an explicit proof of the orthogonality of the bound states 
is missing, both for the Hermitian \cite{dab88}, and the ${\ PT}$ 
symmetric case \cite{ahmed01a}. In the latter case only indirect 
proof has been given for orthogonality of some states \cite{ahmed01a}. 

Obviously, the evaluation of integrals containing the energy 
eigenfunctions of the Scarf II potential is essential to complete 
the analysis of this perfect example for ${\ PT}$ symmetry, 
i.e. to study the behaviour of the pseudonorm and to prove the 
orthogonality of the eigenstates in general. 
In this Letter we present a method to evaluate these integrals 
for the first time, both for the ${\ PT}$ symmetric and the 
Hermitian version of the Scarf II potential, and use these results 
to illustrate the mechanism of ${\ PT}$ symmetry breaking. In 
particular, we prove the orthogonality of the eigenstates and 
derive the normalization constants (the norm) both for the 
${\ PT}$ symmetric and the Hermitian version of this potential, 
and also analyse the ${\ PT}$ symmetric case as a conventional 
complex potential, using the standard (Hilbert space) definition 
of the inner product.

\section{The general form of the Scarf II potential}

Here we follow the notation of Refs. \cite{lgmz00,lgmz01} to 
discuss the Scarf II potential
\begin{eqnarray}
V(x)&=&-\frac{1}{\cosh^2(x)}
\left[\left(\frac{\alpha+\beta}{2}\right)^2
+\left(\frac{\alpha-\beta}{2}\right)^2  -\frac{1}{4}\right]
\nonumber\\
&& +\frac{2{\rm i}\sinh(x)}{\cosh^2(x)}
\left(\frac{\beta+\alpha}{2}\right)
\left(\frac{\beta-\alpha}{2}\right)\ . 
\label{vscarf}
\end{eqnarray}
The bound-state energy eigenvalues are 
\begin{equation}
E^{(\alpha,\beta)}_n=-\left(n+\frac{\alpha+\beta+1}{2}\right)^2\ , 
\label{escarf}
\end{equation}
while the corresponding wavefunctions
\begin{equation}
\psi^{(\alpha,\beta)}_n(x) =C^{(\alpha,\beta)}_n 
(1-{\rm i}\sinh(x))^{\frac{\alpha}{2}+\frac{1}{4}}
 (1+{\rm i}\sinh(x))^{\frac{\beta}{2}+\frac{1}{4}}
P_n^{(\alpha,\beta)}({\rm i}\sinh(x))\ 
\label{fscarf}
\end{equation}
are expressed in terms of Jacobi polynomials \cite{as70} and are 
normalizable if $n<-[{\rm Re}(\alpha+\beta)+1]/2$ holds. 

In the Hermitian case $\alpha$ and $\beta$ are complex and satisfy 
$\alpha^*=\beta$: 
$\alpha=-s-\frac{1}{2}-{\rm i}\lambda$, 
$\beta=-s-\frac{1}{2}+{\rm i}\lambda$ \cite{dab88,lg89}. 
In this case only one regular solution exists. With arbitrary 
$\alpha$ and $\beta$ obviously, the general complex version of the Scarf
II potential is obtained. 

The Scarf II potential can be made ${\ PT}$ symmetric if 
$\alpha^*=\pm\alpha$ and $\beta^*=\pm\beta$ holds \cite{lgmz00}, 
i.e. if $\alpha$ and $\beta$ are both either real or imaginary. 
In order to have real energy eigenvalues both $\alpha$ and $\beta$ 
have to be real, while to have complex bound state spectrum, 
i.e. in the case of spontaneous  breakdown of ${\ PT}$ symmetry 
{\it one of them} has to take an imaginary value \cite{lgmz01}. 
If both $\alpha$ and $\beta$ are imaginary, then there are no 
bound states. Here we assume that $\beta$ is real, and $\alpha$ 
can be real or imaginary, depending on whether the ${\ PT}$ 
symmetry is unbroken or broken. This choice does not restrict the 
generality of the problem, since the roles of $\alpha$ and $\beta$ 
can easily be reversed (see (\ref{fscarfm}) in the Appendix).  

For the Scarf II potential the breakdown of ${\ PT}$ symmetry 
takes place when the strength of the imaginary potential component 
exceeds a certain limit depending on the strength of the real 
potential component, as described in Ref. \cite{ahmed01a}. 
This condition corresponds exactly to taking imaginary values 
for $\alpha$ instead of real ones (see e.g. Ref. \cite{lgmz01} 
for the details), so a smooth transition over the critical point  
can be achieved by moving $\alpha$ to zero along the real axis and 
then continuing along the imaginary axis. 

In the ${\ PT}$ symmetric case there are two sets of normalizable 
solutions \cite{lgmz00,bbcq00,lgfcav01}, which carry the 
upper indexes $(\alpha,\beta)$ and $(-\alpha,\beta)$ in 
(\ref{fscarf}). Obviously, (\ref{vscarf}) is not sensitive to the 
+ or $-$ sign of $\alpha$. In the notation of Ref. \cite{bbcqmz01} 
the two solutions corresponds to quasi-parity $q=+1$ and $-1$. 
This sign difference results in two distinct energy eigenvalues 
in (\ref{escarf}), which form a complex conjugate pair when 
$\alpha$ is imaginary, i.e. in the case of broken ${\ PT}$ 
symmetry. In this case the ${\ PT}$ operation transforms the 
two solutions into each other, while in the unbroken symmetry 
case the two solutions are eigenfunctions of the ${\ PT}$ 
operator. 

In the following sections we are going to evaluate integrals 
containing the standard and ${\ PT}$ symmetric inner product 
of wavefunctions of the type $\psi_n^{(\pm\alpha,\beta)}(x)$ 
(\ref{fscarf}). The technical details of the calculations can be 
found in the Appendix.

\section{The ${\ PT}$ symmetric inner product and the pseudo-norm}
\label{pp}

Let us consider the ${\ PT}$ symmetric inner product 
\cite{mz0103054,japaridze} of two solutions of the type (\ref{fscarf}) 
\begin{equation}
I^{(\alpha,\beta,\delta)}_{nl}=
\int_{-\infty}^{\infty} \psi^{(\alpha,\beta)}_n(x)
[\psi^{(\delta,\beta)}_l(-x)]^* {\rm d}x\ .
\label{ptint}
\end{equation}
According to our choice, $\beta$ is real and $\delta$ can be 
$\pm\alpha$, depending on whether we calculate the ${\ PT}$ 
symmetric inner product of states with the same or different quasi-parities 
($\delta=\alpha$ and $\delta=-\alpha$, respectively), 
furthermore, $\alpha$ can be real or imaginary, depending on 
whether the ${\ PT}$ symmetry is unbroken or broken. 
Applying (\ref{fscarfm}) and (\ref{genint}) presented in the 
Appendix, we can write 
\begin{equation}
I^{(\alpha,\beta,\delta)}_{nl}=
C^{(\alpha,\beta)}_n[C^{(\delta,\beta)}_l]^*
(-1)^n Q^{(\alpha,\beta,\beta,\delta)}_{nl}\ . 
\label{ptintq}
\end{equation}
This formula together with ({\ref{quint}) has significant 
implications regarding the ${\ PT}$ symmetric inner product 
(\ref{ptint}). First note that whenever $\alpha=-\delta^*$ holds, 
the integral vanishes due to the presence of the 
$\sin[\pi(\alpha+\delta^*)]$ term in (\ref{quint}). 
This corresponds to 
either $\alpha=\delta$ with imaginary $\alpha$, i.e. the 
inner product of wavefunctions of the same type (same quasi-parity) 
in the broken ${\ PT}$ symmetry case, or 
$\alpha=-\delta$ with real $\alpha$, i.e. the inner product 
of two different type (different quasi-parity) wavefunctions in the 
unbroken ${\ PT}$ symmetry case. So we can conclude that the 
two states are orthogonal in these situations. 

Now let us consider the cases when $\alpha\ne -\delta^*$. 
The first case is $\delta=\alpha$ with real $\alpha$ (unbroken 
${\ PT}$ symmetry). Substituting this $\delta$ in 
(\ref{quintfin}) (and remembering that $\beta$ is real) we get 
\begin{eqnarray}
I^{(\alpha,\beta,\alpha)}_{nl}&=&\delta_{nl}
\vert C^{(\alpha,\beta)}_n\vert^2 
\frac{2^{\alpha+\beta+2}}{\alpha+\beta+2n+1}
\frac{\sin(\pi\alpha) \sin(\pi\beta)
}{\sin[\pi(\alpha+\beta)]}
\nonumber\\
&&\times
\left(\begin{array}{c} \alpha+\beta+2n \\ n+\beta \end{array}\right)^{-1}
\left(\begin{array}{c} \alpha+\beta+2n \\ n \end{array}\right)\ .
\label{ptintfin}
\end{eqnarray}
This proves directly the orthogonality of the states of the same type 
(i.e. those with the same quasi-parity) for $n\ne l$ when the ${\ PT}$ 
symmetry is unbroken, and gives a closed 
formula for the pseudonorm for $n=l$. Previously this pseudonorm was 
known only for the ground state $n=0$ \cite{bbcqmz01}, while the 
orthogonality of the eigenfunctions was proven only indirectly 
\cite{japaridze,ahmed01a}. This latter proof rests on the equation 
\begin{equation}
(E_n-E_l^*)\int_{-\infty}^{\infty}\psi_n(x)\psi_l^*(-x){\rm d}x=0\ ,
\label{ptortog}
\end{equation}
which is the equivalent of the equation proving the real nature of the 
energy eigenvalues for Hermitan systems. In the case of unbroken 
${\ PT}$ symmetry $E_n$ and $E_l$ are real and they are not equal, 
consequently the integral in (\ref{ptortog}) has to vanish. 

The only remaining case is $\delta=-\alpha$ with imaginary $\alpha$, 
when $\delta^*=\alpha$ holds again. This case gives us the overlap of 
eigenstates belonging to different quasi-parity in the broken 
${\ PT}$ symmetry case. It turns out that the 
$I^{(\alpha,\beta,-\alpha)}_{nl}$ overlap 
has the same form as (\ref{ptintfin}), except that 
$\vert C^{(\alpha,\beta)}_n\vert^2$ has to be replaced with 
$C^{(\alpha,\beta)}_n[C^{(-\alpha,\beta)}_l]^*$. 

Let us summarize the results for the different cases. 
\begin{itemize}
\item
Unbroken ${\ PT}$ symmetry ($\alpha$ real), same type 
wavefunctions: 
$I^{(\alpha,\beta,\alpha)}_{nl}$ is diagonal in $n$ and $l$, 
as seen from Eq. (\ref{ptintfin}). To extract more information, 
we can rewrite Eq. (\ref{ptintfin}) in a somewhat different form, 
after eliminating the sine functions from the formulas by 
combining them with some gamma functions via 
$\Gamma(x)\Gamma(1-x)=\pi/\sin\pi x$: 
\begin{equation}
I^{(\alpha,\beta,\alpha)}_{nl}=\delta_{nl}
(-1)^{n} \pi\vert C^{(\alpha,\beta)}_n\vert^2 
\frac{2^{\alpha+\beta+2}}{(-\alpha-\beta-2n-1)n!}
\frac{\Gamma(-\alpha-\beta-n)}{\Gamma(-\alpha-n)\Gamma(-\beta-n)}\ .
\label{ptintfinalt}
\end{equation}
Due to the condition for having bound states, i.e. 
$n<-[{\rm Re}(\alpha)+\beta+1]/2$, if $\alpha$ is real, then 
in Eq. (\ref{ptintfinalt}) every term is positive, except 
$(-1)^n$ which alternates, and $[\Gamma(-\alpha-n)\Gamma(-\beta-n)]^{-1}$, 
which is real, but its sign depends on the relative magnitude of 
$\alpha$, $\beta$ and $n$. Except for extreme values of $\alpha$ and 
$\beta$ the argument of the two gamma functions is positive for the 
first few $n$'s, so then the alternating $(-1)^n$ factor determines 
the sign of the pseudo-norm, but as $n$ reaches $-\alpha$ 
and/or $-\beta$, this regular pattern changes. The results of this 
case are new, except for $n=0$. 
\item
Unbroken ${\ PT}$ symmetry ($\alpha$ real), different type 
wavefunctions: 
$I^{(\alpha,\beta,-\alpha)}_{nl}=0$,  
due to $\sin\pi(\alpha-\alpha^*)=0$ in (\ref{quint}). 
This was proven indirectly by (\ref{ptortog}) \cite{japaridze,ahmed01a}. 
\item
Broken ${\ PT}$ symmetry ($\alpha$ imaginary), same type 
wavefunctions: 
$I^{(\alpha,\beta,\alpha)}_{nl}=0$,  
due to $\sin\pi(\alpha-\alpha^*)=0$ in (\ref{quint}). 
This was proven indirectly by (\ref{ptortog}) \cite{japaridze,ahmed01a}. 
\item
Broken ${\ PT}$ symmetry ($\alpha$ imaginary), different type 
wavefunctions: 
$I^{(\alpha,\beta,-\alpha)}_{nl}$ is diagonal in $n$ and $l$, 
as seen from Eq. (\ref{ptintfin}). But in this case it seems that 
for $n=l$ there can be two {\it different} wavefunctions which 
are {\bf not} orthogonal, in general. 
Equation (\ref{ptintfinalt}) holds for this case too, except for 
a change in the term containing the normalization constants, as 
discussed before. This non-orthogonality of two different states 
is a new feature of ${\ PT}$ symmetric problems, which in this case 
appears only when the ${\ PT}$ symmetry is broken. 
This unusual result seems to be supported by Eq. (\ref{ptortog}): 
when the ${\ PT}$ symmetry is broken, the energies of the two states 
with the same principal quantum number $n$ but with different 
quasi-parity are complex conjugate to each other, so the zero 
value of (\ref{ptortog}) is secured by the energy term, and the 
integral need not be zero. 
\end{itemize}

\section{The Hermitian Scarf II potential and the normalization 
coefficients of the wavefunctions}
\label{hh}

As discussed in the Introduction, the normalization coefficients 
of the wavefunctions have not been determined as yet, due to 
the involved mathematics \cite{dab88,ahmed01a}. Here we use 
our new approach and show that it can also be applied to evaluate 
integrals for the conventional Scarf II potential. 

Let us now denote the (standard) inner product of the 
states (\ref{fscarf}) as 
\begin{equation}
K^{(\alpha,\beta)}_{nl}=
\int_{-\infty}^{\infty} \psi^{(\alpha,\beta)}_n(x)
[\psi^{(\alpha,\beta)}_l(x)]^* {\rm d}x 
\label{hermint}
\end{equation}
Remember that in this case there is only one set of bound-state 
eigenfunctions, and that in this case we have $\alpha^*=\beta$. 
It is easy to show that 

\begin{equation}
K^{(\alpha,\beta)}_{nl}=
C^{(\alpha,\beta)}_n [C^{(\alpha,\beta)}_l]^* 
Q^{(\alpha,\beta,\alpha,\beta)}_{nl}\ .
\label{hermintq}
\end{equation}
Note that $\delta^*=\alpha^*=\beta$ and $\gamma^*=\beta^*=\alpha$ 
holds now, so the conditions under which the general integral 
can be reduced to a simple form are again satisfied. 
The final result is rather similar to ({\ref{ptintfinalt}) 
obtained for the ${\ PT}$ symmetric case, except that the 
alternating $(-1)^n$ factor is now missing: 
\begin{equation}
K^{(\alpha,\beta)}_{nl}=\delta_{nl}
\pi\vert C^{(\alpha,\beta)}_n\vert^2 
\frac{2^{\alpha+\beta+2}}{(-\alpha-\beta-2n-1)n!}
\frac{\Gamma(-\alpha-\beta-n)}{\Gamma(-\alpha-n)\Gamma(-\beta-n)}\ .
\label{hermintfinalt}
\end{equation}
It is easy to verify that this integral is always positive as it 
should. For this we recall that 
$\alpha^*=\beta$, so $\alpha+\beta$ is real. 
We also recall the condition for bound states
$n<-({\rm Re}[\alpha+\beta]+1)/2$, so the only terms which 
can introduce negative (and complex) quantities are the two 
gamma functions in the denominator. However, their product is 
now $\Gamma(-\alpha-n)[\Gamma(-\alpha-n)]^*
=\vert\Gamma(-\alpha-n)\vert^2>0$, so we conclude that 
$K^{(\alpha,\beta)}_{nn}>0$. 
Equation (\ref{hermintfinalt}) thus determines the normalization 
coefficients of the bound-state wavefunctions of the conventional 
Scarf II potential for the first time: 
\begin{equation}
C^{(\alpha,\beta)}_n=2^{-\frac{\alpha+\beta}{2}-1}
\left[\frac{\Gamma(-\alpha-n)\Gamma(-\beta-n)(-\alpha-\beta-2n-1)n!
}{
\Gamma(-\alpha-\beta-n)\pi
}
\right]^{1/2}\ .
\label{normcont}
\end{equation}

\section{The ${\ PT}$ symmetric Scarf II potential as a 
conventional complex potential}
\label{hp}

As we have discussed in the Introduction, since the ${\ PT}$ 
symmetric Scarf II potential is defined on the real $x$ axis, it 
can also be considered as an ordinary complex potential, and 
the same overlap and normalization integrals can also be evaluated 
for it too. This means using the wavefunctions of the ${\ PT}$ 
symmetric problem in inner products of the type (\ref{hermint}). 
As expected 
for a complex potential, the states cease to be orthogonal, as it 
can be shown by direct calculation of integrals of the type 
(\ref{hermintq}) with real values of $\beta$ and real or imaginary 
values of $\alpha$. Technically this is reflected by the fact 
that the conditions $\delta^*=\alpha$ and $\gamma^*=\beta$, which 
were used in the Appendix to bring the sums to a closed form are 
not satisfied. However, some interesting results can be obtained 
calculating some diagonal matrix elements and overlaps. 

Starting from the one-dimensional Schr\"odinger equation for a 
generic complex potential 
\begin{equation}
-\psi''(x) + [U(x)+{\rm i}W(x)]\psi(x) = (E_R +{\rm i}E_I)\psi(x)
\label{compsch}
\end{equation}
for normalizable eigenfunctions $\psi(x)$, we can derive 
\cite{Ba96,An99} the following 
relation connecting the imaginary component of the potential and 
of the energy eigenvalue: 
\begin{equation}
\frac{
\int_{-\infty}^{\infty} \psi(x)W(x)[\psi(x)]^* {\rm d}x 
}{
\int_{-\infty}^{\infty} \psi(x)[\psi(x)]^* {\rm d}x 
}
=E_I
\label{gal}
\end{equation}
This relation can be rather useful in demonstrating the 
${\ PT}$ symmetry breaking mechanism: tuning the potential 
parameters from the domain of unbroken ${\ PT}$ symmetry 
to the domain of symmetry breaking should result in the 
appearance of a non-zero value of $E_I$. For the Scarf II 
potential this means shifting $\alpha$ along the real axis 
up to zero, and then continuing along the imaginary axis. 

Specifying (\ref{gal}) to (\ref{vscarf}) and  (\ref{fscarf}), 
we get 
\begin{equation}
G^{(\alpha,\beta)}_{n}\equiv
\frac{
\int_{-\infty}^{\infty} \psi^{(\alpha,\beta)}_n(x)
\frac{(\beta^2-\alpha^2)\sinh x}{2\cosh^2 x}
[\psi^{(\alpha,\beta)}_n(x)]^* {\rm d}x 
}{
\int_{-\infty}^{\infty} \psi^{(\alpha,\beta)}_n(x)
[\psi^{(\alpha,\beta)}_n(x)]^* {\rm d}x 
}
\equiv
\frac{
{J}^{(\alpha,\beta)}_{nn}
}{
L^{(\alpha,\beta)}_{nn}
}
\label{galscarf}
\end{equation}
for the integrals, where we assume that $\psi^{(\alpha,\beta)}_n(x)$ 
is a normalizable eigenfunction of the ${\ PT}$ symmetric 
Scarf II potential, i.e. $\beta$ is real, and $\alpha$ is either 
real or imaginary, depending on whether the ${\ PT}$ symmetry is 
unbroken or broken. The integral in the denominator can be expressed 
implicitly in a form similar to the diagonal version of 
Eq. (\ref{ptint}):
\begin{eqnarray}
L^{(\alpha,\beta)}_{nn}&=&
\vert C^{(\alpha,\beta)}_n\vert^2
2^{\frac{\alpha+\alpha^*}{2}+\beta+2}
\frac{\sin[\pi(\alpha^* +\beta)/2] \sin[\pi(\alpha+\beta)/2]
}{\sin[\pi(\beta+(\alpha+\alpha^*)/2)]}
\nonumber\\
&&\times
\sum_{m=0}^n(-1)^m 
\left(\begin{array}{c} n+\alpha \\ m \end{array}\right)
\left(\begin{array}{c} n+\beta \\ n- m \end{array}\right)
\sum_{m'=0}^n (-1)^{m'} 
\left(\begin{array}{c} n+\alpha^* \\ m' \end{array}\right)
\left(\begin{array}{c} n+\beta \\ n- m' \end{array}\right)
\nonumber\\
&&\times
\frac{\Gamma(\frac{\alpha+\beta}{2}+n+1-m+m')
\Gamma(\frac{\alpha^*+\beta}{2}+n+1+m-m')
}{\Gamma(\frac{\alpha+\alpha^*}{2}+\beta+2n+2)}\ .
\label{galintx}
\end{eqnarray}
This integral can also be expressed in terms of the 
$Q^{(\alpha,\beta,\gamma,\delta)}_{nn}$ quantity (\ref{genint}). 
However, in this case we have $\gamma=\alpha$ and $\delta=\beta$, 
so the conditions (\ref{stars}) for reducing the implicit summed 
expression into a closed formula are not satisfied. 
Nevertheless, it is technically not too involved to evaluate 
(\ref{galintx}) for the first few values of $n$. 

A similar expression holds also for the integral appearing in the 
numerator, which can be brought to a sum form using ({\ref{apq1}):  
\begin{eqnarray}
J^{(\alpha,\beta)}_{nn}&=&
\frac{\rm i}{2}(\beta^2-\alpha^2)
\vert C^{(\alpha,\beta)}_n\vert^2
2^{\frac{\alpha+\alpha^*}{2}+\beta}
\frac{\sin[\pi(\alpha^* +\beta^*)/2] \sin[\pi(\alpha+\beta)/2]
}{\sin[\pi(\beta+(\alpha+\alpha^*)/2)]}
\nonumber\\
&&\times
\sum_{m=0}^n(-1)^m 
\left(\begin{array}{c} n+\alpha \\ m \end{array}\right)
\left(\begin{array}{c} n+\beta \\ n- m \end{array}\right)
\nonumber\\
&&\times
\sum_{m'=0}^n (-1)^{m'} 
\left(
\frac{\alpha-\alpha^*}{2}-2m+2m'
\right)
\left(\begin{array}{c} n+\alpha^* \\ m' \end{array}\right)
\left(\begin{array}{c} n+\beta \\ n- m' \end{array}\right)
\nonumber\\
&&\times
\frac{\Gamma(\frac{\alpha+\beta}{2}+n-m+m')
\Gamma(\frac{\alpha^*+\beta}{2}+n+m-m')
}{\Gamma(\frac{\alpha+\alpha^*}{2}+\beta+2n)}\ .
\label{galwintx}
\end{eqnarray}
This expression can also be evaluated directly relatively easily for 
the first few values of $n$. It also turns out that an  
$\alpha-\alpha^*$ can be factored out from the integral 
(\ref{galwintx}). 

From Eq. (\ref{escarf}) we expect 
\begin{equation}
{\rm Im}(E^{(\alpha,\beta)}_n)=
\frac{\rm i}{8}(\alpha-\alpha^*)
(\alpha+\alpha^*+2\beta+4n+2)\ , 
\label{imeptscarf}
\end{equation}
and we indeed get this expression by direct calculation from 
(\ref{galscarf}), (\ref{galintx}) and (\ref{galwintx}). This 
expression vanishes (as it should) for real values of $\alpha$, 
i.e. for unbroken ${\ PT}$ symmetry. When the symmetry is 
broken ($\alpha$ is imaginary), then the second factor in 
(\ref{imeptscarf}) comes from the ratio of the gamma functions 
in the denominator of (\ref{galwintx}) and (\ref{galintx}).

\section{Summary}

We have studied the Scarf II potential in its general form, which 
contains both its Hermitian and its ${\ PT}$ symmetric version, 
depending on the choice of the potential parameters $\alpha$ and 
$\beta$. Our work was motivated by the fact that this potential 
is a perfect laboratory to test the peculiarities of ${\ PT}$ 
symmetric quantum mechanics, and in particular, the mechanism of 
the breakdown of ${\ PT}$ symmetry. This is because it has no 
singularities on the $x$ axis. 

The ${\ PT}$ symmetric version of the Scarf II potential is 
obtained when $\beta$ is chosen to be real, and $\alpha$ real or 
imaginary, depending on whether the ${\ PT}$ symmetry is broken 
or unbroken: in the former case the bound-state energy spectrum is 
completely real, while in the latter case it is fully complex. 
The reality of $\beta$ is not an essential restriction, 
since the roles of $\alpha$ and $\beta$ can be interchanged in a 
trivial way. The potential is not sensitive to the sign of 
$\alpha$, and this is reflected by the fact that both in the 
unbroken and broken ${\ PT}$ symmetry case there are two sets 
of normalizable (bound) states with the same value of the principal 
quantum number $n$, and they are distinguished by the sign of $\alpha$, 
i.e. the quasi-parity. The corresponding energy eigenvalues are 
different: they are both real for unbroken ${\ PT}$ symmetry, 
and they form a complex conjugate pair when the symmetry is broken. 

We have devised a way by which the modified inner product of 
the wavefunctions 
$I^{(\alpha,\beta,\pm\alpha)}_{nl}\equiv\langle \psi^{(\alpha,\beta)}_n
\vert {\ P} \vert \psi^{(\pm\alpha,\beta)}_l\rangle$ 
can be evaluated explicitly. We have shown that these states 
are orthogonal in $n$ and in the quasi-parity quantum numbers, 
except when $\alpha$ is imaginary, i.e. the ${\ PT}$ symmetry is 
broken, and the inner product of two states with the same $n$ but 
different quasi-parities is considered. We derived a closed 
expression for the diagonal inner products, i.e. the pseudo-norm, 
and we found that its sign is indefinite, as expected from more 
general considerations. However, the sign of the pseudo-norm 
depends on $\alpha$ and $\beta$ in a rather complicated way, so 
no compact rule can be formulated for its sign. 

Using the same techniques we also determined the standard inner product 
of the Hermitian Scarf II wavefunctions. In this case 
$\alpha^*=\beta$ holds, and there is only one set of bound-state 
wavefunctions; the other one represents resonances: both  
the orthogonality of the states, the normalization constants 
for the Sacrf II potential have been derived for the first time. 

We also analysed the ${\ PT}$ symmetric Scarf II potential as 
an ordinary complex potential. For this, we evaluated two types of 
diagonal matrix elements (using the standard inner product) of the 
${\ PT}$ symmetric wavefunctions: the matrix element of the 
imaginary component of the potential and the norm. The ratio of 
these gives a closed expression for the imaginaly component of the 
energy, which contains the potential parameters and thus can be 
used to study the breakdown of ${\ PT}$ symmetry as the value 
of $\alpha$ is tuned from real values to imaginary ones through 0. 
In the case of other solvable ${\ PT}$ symmetric potentials 
this transition requires crossing a singular point of the potential, 
which obscures important aspects of ${\ PT}$ symmetry breaking.

\section*{Acknowledgment}
This work was supported by the OTKA grant No. T031945 (Hungary), 
INFN and ENEA (Italy).

\section*{Appendix}

Here we derive a general formula for the integrals which includes as 
special cases both the Hermitian and the ${\ PT}$ symmetric 
inner product of the wavefunctions. 

First, let's define the functions
\begin{equation}
F^{(\alpha,\beta)}_n(x) =
(1-{\rm i}\sinh(x))^{\frac{\alpha}{2}+\frac{1}{4}}
 (1+{\rm i}\sinh(x))^{\frac{\beta}{2}+\frac{1}{4}}
P_n^{(\alpha,\beta)}({\rm i}\sinh(x))\ , 
\label{fgenscarf}
\end{equation}
which are the bound-state Scarf II wavefunctions (\ref{fscarf}) 
without the normalization constant. 
The Jacobi polynomials appearing in (\ref{fgenscarf}) take the form 
\cite{as70}
\begin{eqnarray}
P_n^{(\alpha,\beta)}({\rm i}\sinh(x))&=&
\frac{1}{2^n}\sum_{m=0}^n
\left(\begin{array}{c} n+\alpha \\ m \end{array}\right)
\left(\begin{array}{c} n+\beta \\ n- m \end{array}\right)
\nonumber\\
&&\times
(-1)^{n-m}(1-{\rm i}\sinh(x))^{n-m}(1+{\rm i}\sinh(x))^{m}\ .
\label{jacobi}
\end{eqnarray}
Also due to the properties of the Jacobi polynomials \cite{as70} 
the (\ref{fgenscarf}) functions transform under spatial reflection 
(i.e. the ${\ P}$ operation) in the following way:
\begin{eqnarray}
F^{(\alpha,\beta)}_n(-x) &=& 
(1+{\rm i}\sinh(x))^{\frac{\alpha}{2}+\frac{1}{4}}
 (1-{\rm i}\sinh(x))^{\frac{\beta}{2}+\frac{1}{4}}
P_n^{(\alpha,\beta)}(-{\rm i}\sinh(x))
\nonumber\\
&=&
(-1)^n 
(1-{\rm i}\sinh(x))^{\frac{\beta}{2}+\frac{1}{4}}
(1+{\rm i}\sinh(x))^{\frac{\alpha}{2}+\frac{1}{4}}
P_n^{(\beta,\alpha)}({\rm i}\sinh(x))\ .
\nonumber \\
&=& (-1)^nF^{(\beta,\alpha)}_n(x) 
\label{fscarfm}
\end{eqnarray}

Now define the integral 
\begin{equation}
Q^{(\alpha,\beta,\gamma,\delta)}_{nl}=
\int_{-\infty}^{\infty} F^{(\alpha,\beta)}_n(x)
[F^{(\gamma,\delta)}_l(x)]^* {\rm d}x \ .
\label{genint}
\end{equation}
This integral contains the formulae necessary to evaluate the 
standard
(Hermitian) inner product of the ordinary (real) Scarf II bound-state 
eigenfunctions (with $\gamma=\alpha$ and $\delta=\beta$) and 
the ${\ PT}$ symmetric inner product of the eigenfunctions 
of the ${\ PT}$ symmetric Scarf potential with equal and 
different quasi-parity (with $\gamma=\beta$, $\delta=\alpha$ and 
$-\alpha$, respectively). Note that in the latter case the roles 
of $\gamma$ and $\delta$ are interchanged, due to (\ref{fscarfm}). 
The above formula can also be used, of course, to evaluate integrals 
appearing in the Hermitian inner product of the eigenfunctions of 
the ${\ PT}$ symmetric Scarf II potential, however, we shall find 
that in this case it cannot be reduced to simpler forms. 

First we need to evaluate integrals of the kind 
\begin{equation}
A^{(p,q)}_0\equiv\int_{-\infty}^{\infty}
(1-{\rm i}\sinh x)^p(1+{\rm i}\sinh x)^q{\rm d} x\ , 
\label{apq0}
\end{equation}
which then appear in sums in (\ref{genint}). The integral can then 
be evaluated in a multistep procedure \cite{bbcqmz01,prud}:
\begin{eqnarray}
A^{(p,q)}_0&=&\int_{-\infty}^{\infty}
\cosh^{p+q}x \exp[(q-p)\tanh^{-1}({\rm i}\sinh x)]
\label{apq0a}
\\
&=&\int_{-\infty}^{\infty}
\cosh^{p+q}x \exp[{\rm i}(q-p)\tan^{-1}(\sinh x)]
\label{apq0b}
\\
&=&\int_{-\pi/2}^{\pi/2}
\cos^{-(p+q+1)}y \exp[{\rm i}(q-p)y]
\label{apq0c}
\\
&=&2^{p+q+1}\pi\frac{\Gamma(-p-q)}{\Gamma(-p+1/2)\Gamma(-q+1/2)}\ .
\label{apq0d}
\end{eqnarray}
In (\ref{apq0a}) we applied the relation 
$\tanh^{-1} z = \frac{1}{2}\ln\frac{1+z}{1-z}$ 
which is also used sometimes to write the Scarf II wavefunctions 
in an alternative form \cite{dab88,ahmed01a,bbcqmz01}, and in 
(\ref{apq0c}) a change of variables was performed: 
$\tan y=\sinh x$. This method was used in Ref. \cite{bbcqmz01} 
to evaluate the pseudo-norm of the ground-state wavefunctions for 
unbroken ${\ PT}$ symmetry. In what follows we shall also find it 
useful to evaluate another integral, which follows directly from 
(\ref{apq0d}): 
\begin{equation}
A^{(p,q)}_1\equiv\int_{-\infty}^{\infty}
\sinh x (1-{\rm i}\sinh x)^p(1+{\rm i}\sinh x)^q{\rm d} x 
={\rm i}\frac{p-q}{p+q+1}A^{(p,q)}_0\ .
\label{apq1}
\end{equation}

In the next step we have to evaluate sums of integrals of the type 
(\ref{apq0}) appearing in (\ref{genint}): 
\begin{eqnarray}
Q^{(\alpha,\beta,\gamma,\delta)}_{nl}&=&
(-1)^{n+l}
2^{\frac{\alpha+\beta+\gamma^*+\delta^*}{2}+2}
\frac{\sin[\pi(\alpha+\delta^*)/2] \sin[\pi(\beta+\gamma^*)]
}{\sin[\pi(\alpha+\beta+\gamma^*+\delta^*)/2)]}
\nonumber\\
&&\times
\sum_{m=0}^n(-1)^m 
\left(\begin{array}{c} n+\alpha \\ m \end{array}\right)
\left(\begin{array}{c} n+\beta \\ n- m \end{array}\right)
\sum_{m'=0}^l (-1)^{m'} 
\left(\begin{array}{c} l+\gamma^* \\ m' \end{array}\right)
\left(\begin{array}{c} l+\delta^* \\ l- m' \end{array}\right)
\nonumber\\
&&\times
\frac{\Gamma(\frac{\alpha+\delta^*}{2}+n-m+m'+1)
\Gamma(\frac{\beta+\gamma^*}{2}+l+m-m'+1)
}{\Gamma(\frac{\alpha+\beta+\gamma^*+\delta^*}{2}+n+l+2)}\ .
\label{quint}
\end{eqnarray}
Without the loss of generality we can assume that $n\le l$. 
With some rearrangement of the binomial coefficient and the 
gamma functions we can rewrite the last sum (over $m'$) into 
\begin{eqnarray}
&&\frac{\Gamma(l+\gamma^*+1)\Gamma(l+\delta^*)}{
\Gamma(\frac{\alpha+\beta+\gamma^*+\delta^*}{2}+n+l+1)l!}
\sum_{m'=0}^l(-1)^{m'} 
\left(\begin{array}{c} l \\ m' \end{array}\right)
\nonumber\\
&&\times 
\frac{\Gamma(\frac{\alpha+\delta^*}{2}+n-m+m'+1)}{\Gamma(\delta^*+m'+1)}
\frac{\Gamma(\frac{\beta+\gamma^*}{2}+l+m-m'+1)}{\Gamma(\gamma^*+l-m'+1)}
\ .
\label{sumxw}
\end{eqnarray}
In what follows we assume that the relations 
\begin{equation}
\delta^*=\alpha \hskip 1cm \gamma^*=\beta
\label{stars}
\end{equation}
hold, in order to bring (\ref{sumxw}) to a simpler form. 
This assumption turns out to be valid for the most important 
cases, as it is described in the appropriate sections. 
Under the conditions (\ref{stars}) the two terms containing the 
ratios of gamma functions in (\ref{sumxw}) can be 
written as a finite power series of $m'$: $\sum_{j=0}^n c_j (m')^j$. 
This clearly follows from the two ratios of the gamma functions: the 
first one contains $m'$ up to the $(n-m)$'th power in the form 
$(\alpha+m'+1)(\alpha+m'+2)\dots(\alpha+m'+n-m)$, 
while the second one contains it up to the $m$'th power as 
$(\beta+l-m'+1)(\beta+l-m'+2)\dots(\beta+l-m'+m)$. 
Now observing Eqs. 4.2.2.3 and 4.2.2.4 of Ref. \cite{prud} 
we find that 
\begin{eqnarray}
\sum_{m'=0}^l(-1)^{m'} 
\left(\begin{array}{c} l \\ m' \end{array}\right)
\sum_{j=0}^n c_j (m')^j 
&=&
\sum_{j=0}^n c_j 
\sum_{m'=0}^l(-1)^{m'} 
\left(\begin{array}{c} l \\ m' \end{array}\right)
(m')^j
\nonumber\\
&&
=\sum_{j=0}^n c_j \delta_{jl}(-1)^l l!
\label{prud422}
\end{eqnarray}
Since we assumed that $n\le l$ holds, $j\le l$ is also valid, so 
this sum can be non-zero only for $n=l$. In this case its value is 
\begin{equation}
\delta_{ln}(-1)^n n! c_n = 
\delta_{ln}(-1)^n n! (-1)^m = 
\delta_{ln} n! (-1)^{n+m} \ . 
\label{prudx}
\end{equation}
Here we used that $c_n$, the coefficient of $(m')^n$ 
(the highest possible power of $m'$) in the power series 
is $(+1)^{n-m} (-1)^m$, as can be seen from the factorization of the 
ratio of the gamma functions above. Substituting this result into 
(\ref{sumxw}) and (\ref{quint}) and then using Eq. 4.2.5.19 of 
Ref. \cite{prud}, we get 
\begin{eqnarray}
Q^{(\alpha,\beta,\gamma=\beta^*,\delta=\alpha^*)}_{nl}&=&
\delta_{nl}(-1)^n
\frac{2^{\alpha+\beta+2}}{\alpha+\beta+2n+1}
\frac{\sin(\pi\alpha) \sin(\pi\beta)
}{\sin[\pi(\alpha+\beta)]}
\nonumber\\
&&\times\left(\begin{array}{c} 
\alpha+\beta+2n \\ n+\beta \end{array}\right)^{-1}
\left(\begin{array}{c} \alpha+\beta+2n \\ n \end{array}\right)\ .
\label{quintfin}
\end{eqnarray}

Finally, we note that for the $n=l$ diagonal case (\ref{quint}) 
can be evaluated in an alternative way too, using the sum rule 
\begin{equation}
\sum_{k=0}^n (-1)^k
\left(\begin{array}{c} n \\ k \end{array}\right)
\left(\begin{array}{c} a-m-k \\ n-m \end{array}\right)
\left(\begin{array}{c} b+m+k \\ m \end{array}\right)
=(-1)^m
\left(\begin{array}{c} n \\ m \end{array}\right)\ .
\label{alter}
\end{equation}
The interesting feature of this result is that the right hand side 
is independent of $a$ and $b$. This formula is missing from 
the standard compilation \cite{prud}, and can be proven by 
an induction in $n$, using also the properties of the binomial 
coefficients.

\end{document}